\documentstyle[sprocl,psfig]{article}

\def\PRD#1{{\em Phys. Rev.} {\bf D#1}}

\def\al{\alpha}

\def\be{\begin{equation}}
\def\ee{\end{equation}}
\def\bea{\begin{eqnarray}}
\def\eea{\end{eqnarray}}
\def\bean{\begin{eqnarray*}}
\def\eean{\end{eqnarray*}}
\def\half{\frac{1}{2}}
\def\dslash{/\!\!\! \partial}

\def\Aslash{/\!\!\!\! A}
\def\gm{\gamma}
\def\bra{\langle}
\def\ket{\rangle}

\begin{document}

\title{DYNAMICS OF FERMIONS AND INHOMOGENEOUS BOSE FIELDS ON A REAL-TIME 
LATTICE\footnote{Talk presented by G.~Aarts
at Strong and Electroweak Matter '98, Kopenhagen, Denmark, December 2-5, 
1998.} 
}

\author{GERT AARTS${}^1$ and JAN SMIT${}^{1,2}$}

\address{${}^1$Institute for Theoretical Physics, Utrecht University\\
Princetonplein 5, 3584 CC Utrecht, the Netherlands\\
${}^2$Institute for Theoretical Physics, University of Amsterdam\\
Valckenierstraat 65, 1018 XE Amsterdam, the Netherlands
}

\maketitle
\abstracts{
The dynamics of the $1+1$ abelian Higgs model with fermions is 
studied in the large $N_f$ approximation, on a real-time lattice. The Bose 
fields obey effective classical equations of motion which include the 
fermion back reaction. The dynamics of the quantized fermion field is 
treated with a mode function expansion. Numerical results are shown for 
renormalizability, nonequilibrium dynamics and the anomalous charge, and 
Pauli blocking. }

\section{Introduction}

Non-perturbative approximations to real-time dynamics in Quantum Field 
Theory include large $N$, Hartree or semiclassical 
approximations.\cite{largeN,applic} Fields are written as a sum of a mean 
or classical field and a quantum part (treated in practice with a 
mode function expansion), and effective equations of motion can be 
derived that couple them. 
Such an approximation includes e.g. the Hard Thermal Loops in $3+1$ 
dimensional hot gauge theories. Furthermore, the effective equations 
contain the divergences of the quantum theory, and these can be 
renormalized in the usual way.

However, in actual numerical implementations, the emphasis has 
been on homogeneous mean fields.\cite{applic} This excludes dynamics 
related to inhomogeneous classical configurations such as kinks, 
sphalerons, and 
bubbles. Furthermore, when the quantum mode functions interact 
via a homogeneous mean field, there is no `mode mixing'.  
This makes the long-time behaviour of the system peculiar.
Allowing for inhomogeneous mean fields immediately removes the last two 
objections. The drawback is that the numerical problem becomes much more 
demanding. Space dependence of the mode functions is no longer given by 
plane waves, but is determined by the full partial differential equations. 
Here we'll discuss results for a $1+1$ dimensional model, where the 
numerics can be handled with brute force. A more detailed account, and much 
more technical details, can be found elsewhere.\cite{AaSm98}
	
\section{Abelian Higgs model with fermions and effective equations of 
motion}

A toy model for electroweak baryogenesis is the $1+1$ dimensional abelian 
Higgs model with axially coupled fermions. The action is given by
\bean
S = -\int d^2 x \Big[\frac{1}{4e^2}F_{\mu\nu}F^{\mu\nu}
   + |D_\mu \phi|^2 +
\lambda(|\phi|^2 - \half v^2)^2
&&\\
 + \bar\psi(\dslash + \half i\Aslash\gm_5)\psi + G
\bar\psi(\phi^*P_L+\phi P_R)\psi\Big].
&&
\eean
As the electroweak theory, this model has a non-trivial 
bosonic vacuum, sphalerons, and an anomalous global symmetry: 
fermion number violation.

We want to study the dynamics numerically in real-time. We find it then 
convenient to use a real-time lattice formulation, for several reasons: the 
lattice acts as a gauge invariant ultraviolet cutoff, there are exact 
symmetries for finite lattice spacing, and the resulting integration 
algorithm is simple and stable.
Fermions on a lattice give of course rise to the fermion doubler problem. 
To deal with this, we use Wilson's fermion method in space, and interpret 
the doubler in time as a second flavour. 
For technical reasons, it is convenient to transform the model to one 
with a vector gauge symmetry, by performing charge conjugation on the 
right-handed fermions only. The anomalous current is then axial, with 
the anomaly equation $Q_5(t)-Q_5(0) = C(t)-C(0)$, where $C=-\sum_x A_1/2\pi$ 
is the Chern-Simons number. 
Finally, it is also convenient to use a real 4-component Majorana field 
$\Psi$.

Effective equations of motion can be derived from first principles, by 
duplicating the fermion field $N_f$ times, and taking $N_f\to 
\infty$.\cite{largeN,AaSm98} In the language of the Introduction, the Bose 
fields represent the mean fields, and the fermions the fluctuations. 
The resulting bosonic equations are (in the temporal gauge $A_0=0$)
\bea
&&\partial'_0\partial_0 A_{1} = e^2 (j_{h}^1 + \bra
j_{f}^1\ket),\;\;\;\;\;\;\;\;
\partial_1'\partial_0 A_{1} = -e^2 (j_{h}^0 + \bra
j_{f}^0\ket),\\ &&
\label{eqscalar}
\partial_0'\partial_0 \phi = D'_1 D_1\phi - 2\lambda
(|\phi|^2 - \half v^2_B)\phi + G\bra F\ket.
\eea
These are similar to the classical equations, but with the fermion 
back reaction present, i.e. the fermion current $\bra 
j_f^\mu\ket=\frac{i}{4}\bra\Psi^T\beta\gm^\mu\rho_2\Psi\ket$ and 
force $\bra F\ket=\frac{i}{4}\bra\Psi\beta(\rho_1+i\rho_3)\Psi\ket$ (in 
continuum notation). $\rho_i$ are matrices appearing in the Majorana 
description. The fermions 
are treated with a mode function expansion, which in the continuum would read
\[ \Psi(x,t) = \sum_\al \left[ b_\al U_\al(x,t) + b_\al^\dagger 
U_\al^*(x,t)\right].\]
$\al$ labels the complete set of mode functions (it typically contains 
momentum), the 
creation/annihilation operators are time independent and determine the 
initial quantum state (we use vacuum, $\bra b_\alpha^\dagger 
b_\alpha\ket=0$), and the mode functions themselves are solutions of 
the Dirac equation
\be  \half i(\partial_0+\partial_0')U_\al(x,t) = {\cal 
H}_D(A,\phi)U_\al(x,t),
\ee
with ${\cal H}_D(A,\phi)$ the Dirac hamiltonian in presence of the Bose 
fields.
Note that we now have a closed set of equations.
For a precise treatment of the lattice fermion doublers in time, and the 
initial conditions for the mode functions, we refer to our 
paper.\cite{AaSm98}
The initial conditions for the Bose fields are such that only long wave 
lengths are excited, and that Gauss' law (1b) is satisfied.

\section{Renormalizability, nonequilibrium dynamics, and Pauli blocking}

\begin{figure}
\centerline{
\psfig{figure=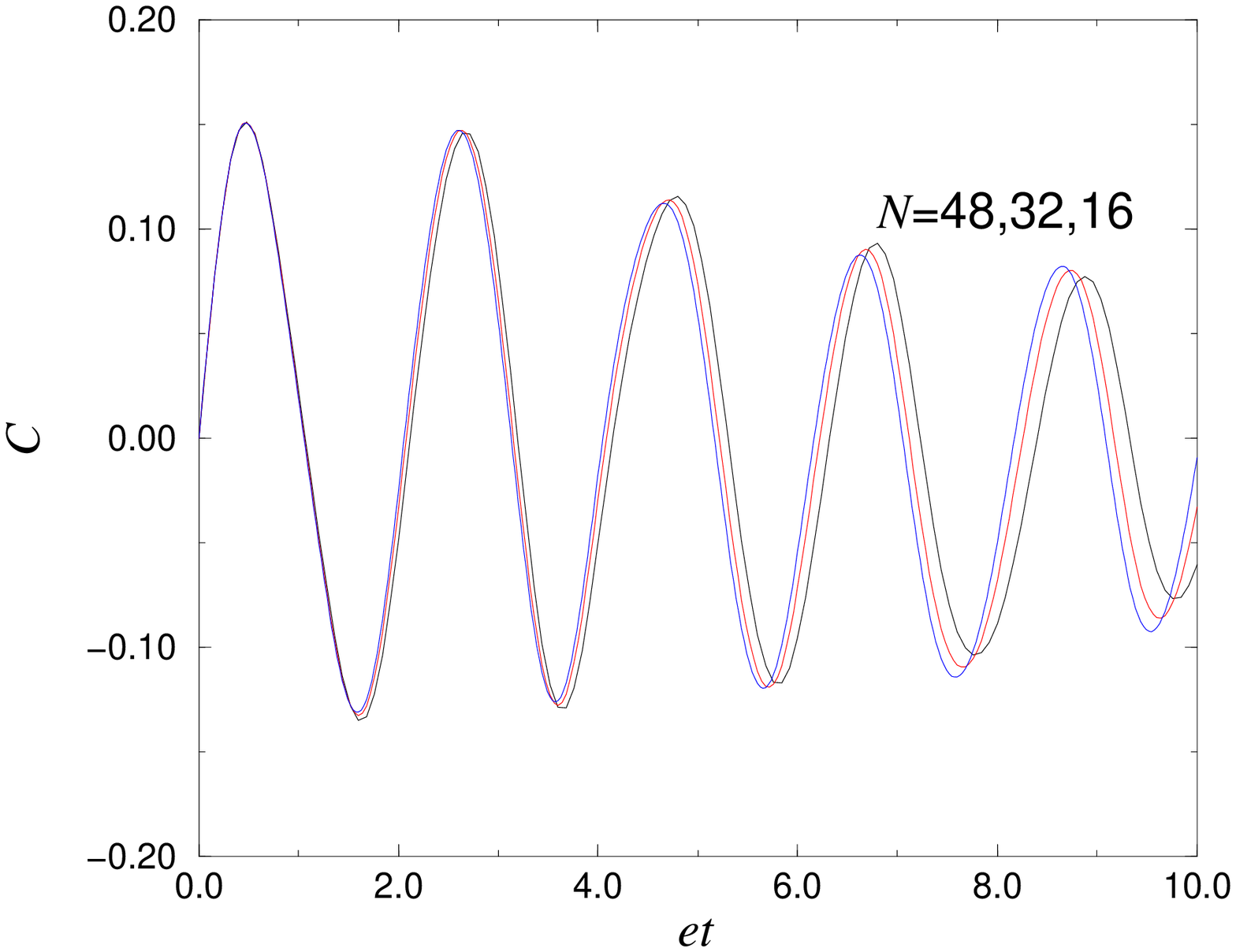,height=5.2cm}
\hspace{-0.5cm}
\psfig{figure=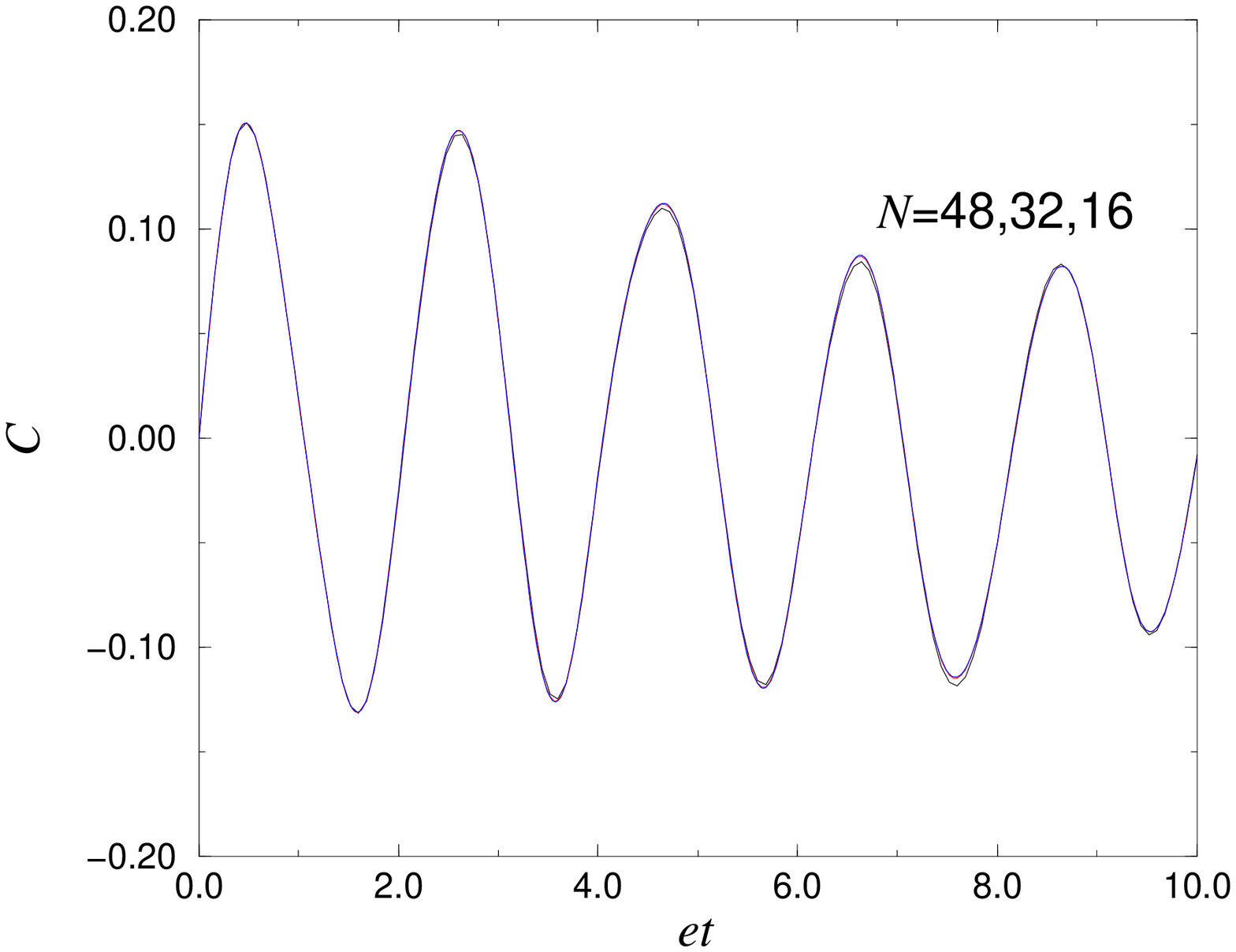,height=5.2cm}
}
\vspace{-0.5cm}
\caption{Chern-Simons number $C$ for three values of the 
cutoff, $N = 16, 32, 48$ and 
G/e = 0.5. Left: the bare $v_B^2=10$ is kept fixed. Right: the renormalized 
$v_R^2 = 11.25$ is kept fixed.} 
\label{figrenorm}
\end{figure}

The equations contain the ultraviolet divergences of the quantum theory, 
and need to be renormalized.  
In the full quantum theory only the scalar self energy is divergent, and 
because we only integrated out the fermions, we expect to find only the 
fermion loop contribution. 
Let's take a closer look to the scalar field equation 
(\ref{eqscalar}). If we take for simplicity $\phi=v_R/\sqrt{2}$ (i.e. 
equal to the renormalized v.e.v., $v_B$ is the bare parameter), and $A_1=0$, 
and also evaluate the fermions in 
this background, (\ref{eqscalar}) reduces to the gap equation 
\be
\label{eqgap}
\lambda(v_R^2-v_B^2)\frac{v_R }{\sqrt{2}} = G\bra F\ket (v_R), 
\;\;\;\;
\bra F\ket(v_R) = \frac{Gv_R}{\sqrt{2}}\left(\frac{1}{\pi}\log N + {\cal 
O}(N_0)\right).\ee
An explicit calculation of $\bra F\ket$ shows that it is indeed a 
logarithmically 
divergent sum, indicated with $\log N$ in (\ref{eqgap}b) ($N$ is the 
number of spatial lattice points). This divergence is canceled by the 
appropriate $v_B$. In 
practice, we fix $v_R$, and then find (for certain $N$, i.e. lattice 
spacing) the corresponding bare parameter $v_B$, from  (\ref{eqgap}a).
The result of this procedure is demonstrated in fig.~\ref{figrenorm}.
As expected, a proper renormalization gives converging physical 
results.

\begin{figure}
\centerline{
\psfig{figure=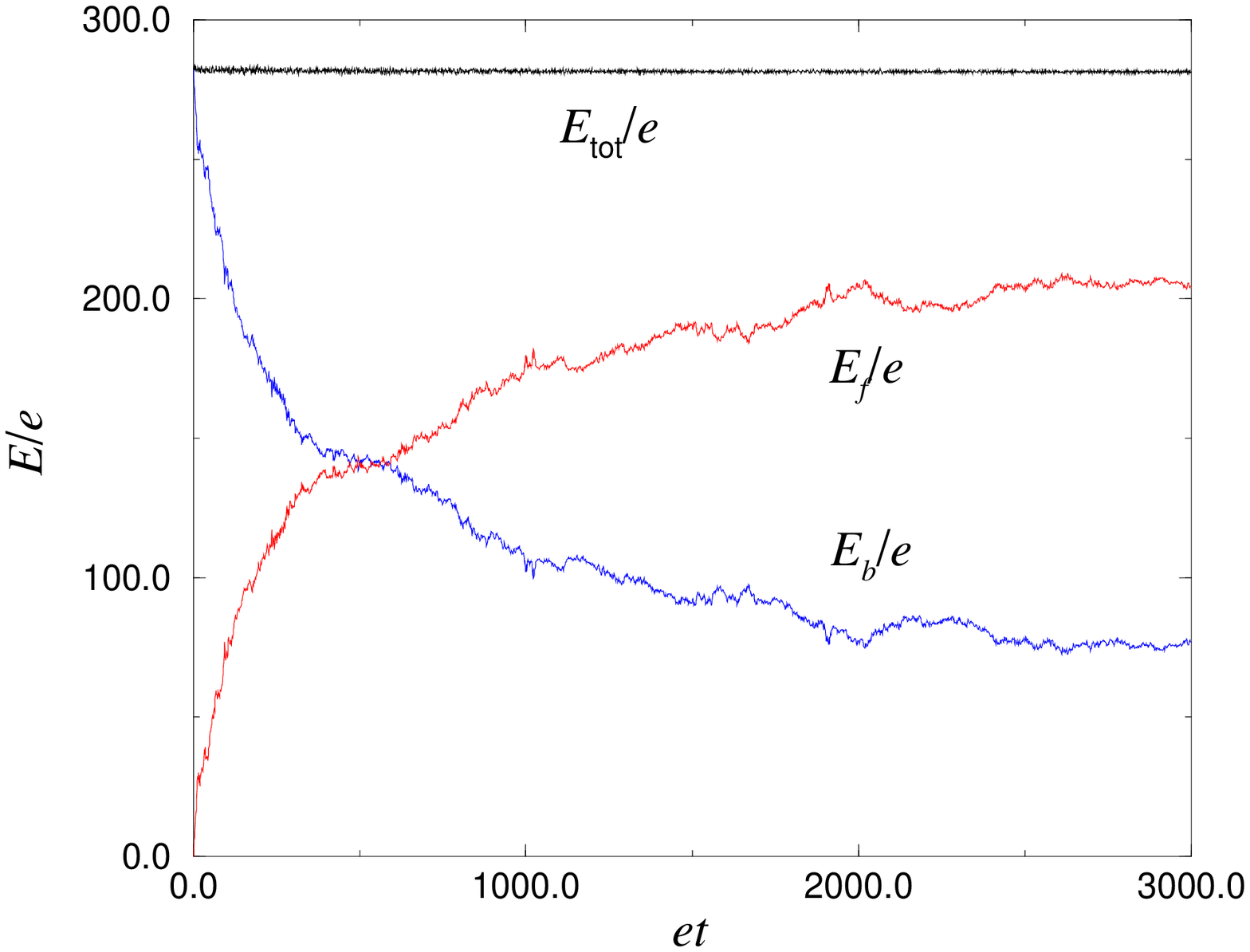,height=5.2cm}
\hspace{-0.5cm}
\psfig{figure=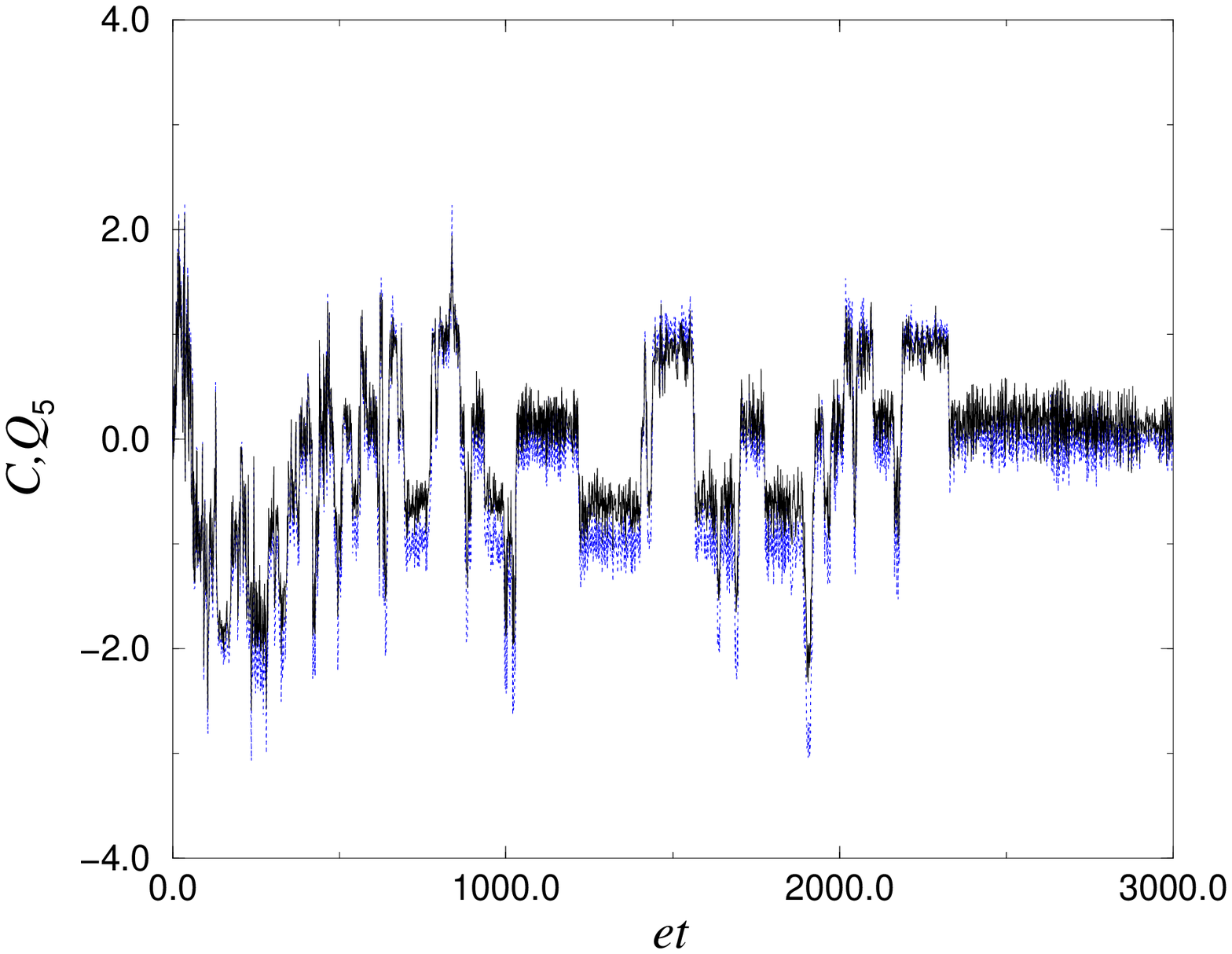,height=5.2cm}
}
\vspace{-0.5cm}
\caption{
Left: Energy of the Bose fields, fermion field, and the conserved sum. 
Right: Chern-Simons number (dashed), and the anomalous charge (solid), 
for $G/e=0$.} \label{fignonpert} 
\end{figure}

In fig.~\ref{figrenorm}, the Chern-Simons number undergoes a damped 
oscillation, with a small amplitude. Remember that the sphaleron 
sits at half-integer $C$. To probe this non-perturbative physics, we put 
more energy in the initial Bose fields. The results are shown in 
fig.~\ref{fignonpert}. The initial state is clearly 
out-of-equilibrium and there is energy transfer from the Bose fields to 
the fermions. The Chern-Simons number is initially very wild, but as the 
bosonic energy decreases, the Bose fields start to feel the sphaleron 
barriers, and plateaus become visible. The anomalous charge follows the 
Chern-Simons number, in accordance with the anomaly equation: 
$\Delta Q_5 = \Delta C$.

\begin{figure}
\centerline{
\psfig{figure=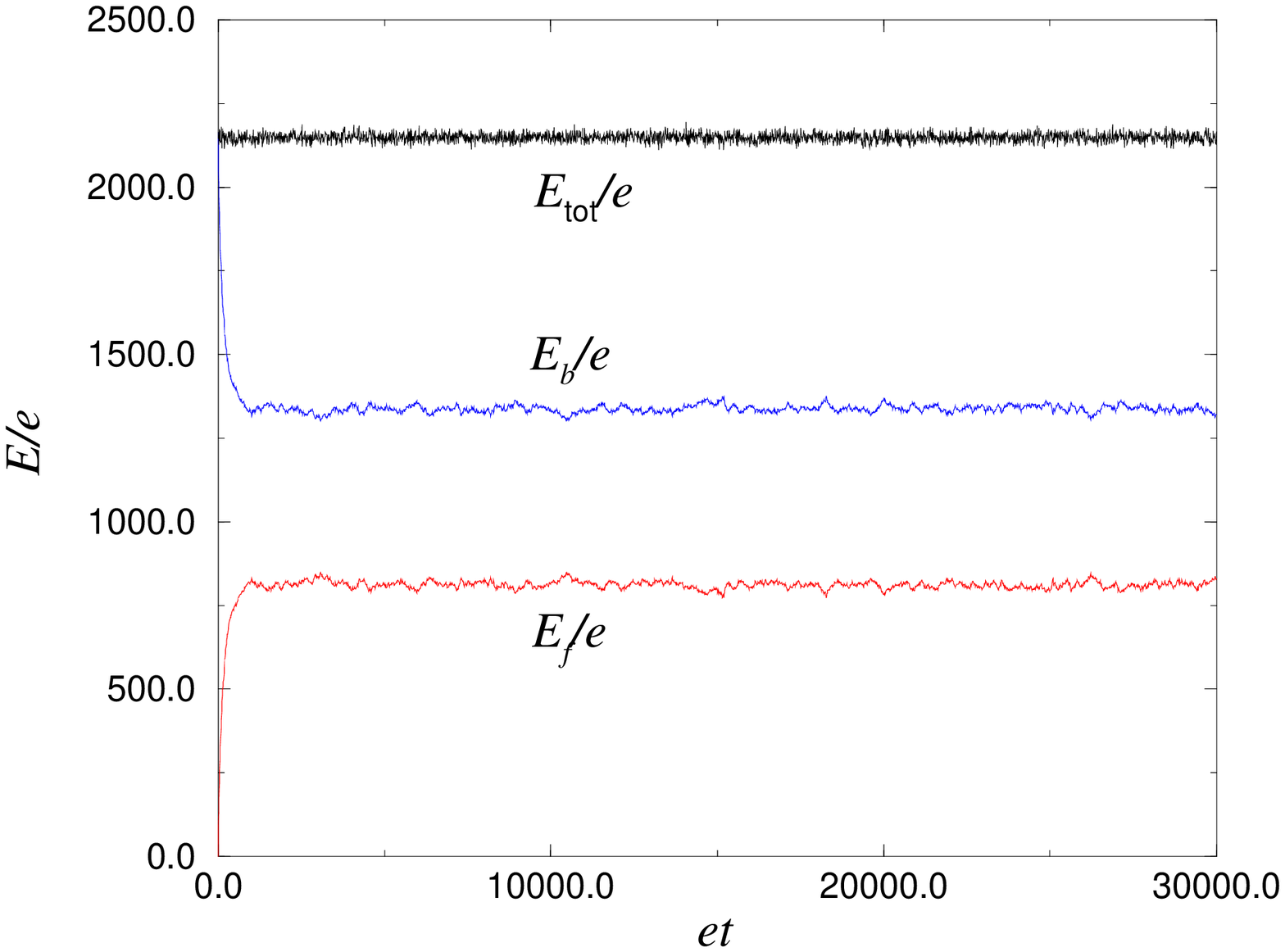,height=5.2cm}
\hspace{-0.5cm}
\psfig{figure=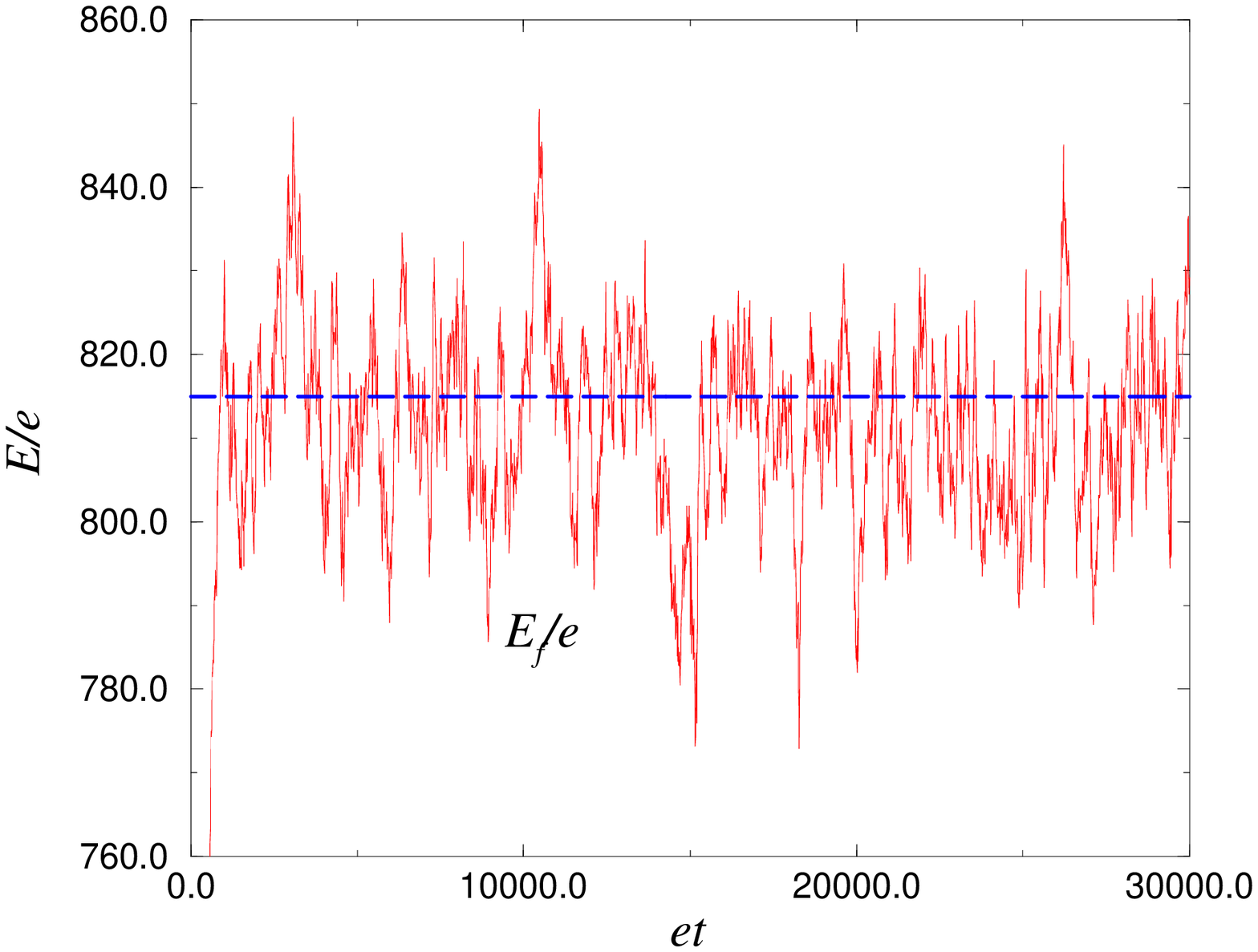,height=5.2cm}
}
\vspace{-0.5cm}
\caption{Left: as in fig.~\ref{fignonpert} (left), with $G/e=0.1$.
Right: blowup of the fermion energy, the `free' maximal fermion energy is 
indicated with a dashed line.}
\label{figpauli}
\end{figure}

Energy transfer from the Bose fields to the fermionic degrees of freedom 
cannot go on forever because of Pauli blocking. 
This becomes clear if we choose the initial energy in the Bose fields  
to be much larger than before (compare 
figs.~\ref{fignonpert},~\ref{figpauli}). Again there is rapid energy 
transfer (until approx. $et=1000$), but then there is a rather sharp 
transition after which 
the energy in both subsystems remains approximately constant. A heuristic 
picture is as follows: the state with maximal 
fermion energy is given by a completely filled Dirac sea {\em and} sky. 
Note that this situation is not physical, it 
is a manifestation of the fact that for finite $N$ we only have a finite 
number of states.  Nevertheless, a calculation of the `free' fermion energy 
(i.e. neglecting the fluctuating Bose fields, keeping only 
$\phi=v_R/\sqrt{2}$) when all states in the spectrum are completely 
filled, gives for our choice of parameters and for $N=32$, $E_f^{\rm 
max}/e = 815$. Indeed, a blowup shows that the fermion energy fluctuates 
around this value. 

\section{Summary}
As an approximation to non-perturbative dynamics in quantum field theory, we 
studied a coupled system of classical Bose fields and a quantized fermion 
field. Skipping many technical details, we presented three numerical 
results. 
Many more questions can be addressed in this framework (e.g. long time 
behaviour, inclusion of CP violation, sphaleron 
rate in the presence of fermions), and we hope to report on these issues 
in the future.

\section*{Acknowledgments}
This work is supported by FOM.


\section*{References}
\begin{thebibliography}{99}
\bibitem{largeN}
        F.~Cooper, S.~Habib, Y.~Kluger, E.~Mottola, J.P. Paz and
        P.R. Anderson, \PRD{50} (1994) 2848.
\bibitem{applic}
	See e.g. 
        F.~Cooper, S.~Habib, Y.~Kluger and E.~Mottola,
        \PRD{55} (1997) 6471;
	D.~Boyanovsky et.~al., \PRD{57} (1998) 2166; 
        J.~Baacke, K.~Heitmann and C.~P\"atzold, \PRD{58} (1998) 125013; 
        and references therein.
\bibitem{AaSm98} G.~Aarts and J.~Smit, {\tt hep-ph/9812413}.


\end{thebibliography}
\end{document}